\documentclass[epj]{webofcm}
\usepackage[varg]{txfonts}
\usepackage{color}
\wocname{}
\woctitle{}

\newcommand{\DP}{\Delta\Pi}
\newcommand{\ind}[2]{^{\mbox{\scriptsize $#1$}}_{\mbox{\scriptsize #2}}}
\newcommand{\inds}[2]{^{\mbox{\scriptsize $#1$}}_{\mbox{\tiny #2}}}

\newcommand{\nf}{n_{\mbox{\scriptsize f}}}

\newcommand\AL[1]{\Delta\alpha\ind{#1}{lep}}
\newcommand\AH[1]{\Delta\alpha\ind{#1}{had}}
\newcommand\AMH[1]{\Delta\alpha\ind{#1}{had}}
\newcommand{\eqnsgn}[1]{\!\!\!\!& #1 &\!\!\!\!}
\newcommand{\txt}[2]{\color{#2}\scriptsize\textsf{#1}}
\definecolor{G4}{rgb}{0.00,0.55,0.00}
\definecolor{RB4}{rgb}{0.15,0.25,0.55}

\begin{document}

\selectlanguage{english}
\title{Dispersive approach to QCD and hadronic contributions \\ to electroweak observables}

\author{Alexander V.~Nesterenko\inst{1}\fnsep\thanks{\email{nesterav@theor.jinr.ru}}}

\institute{Bogoliubov Laboratory of Theoretical Physics,
Joint Institute for Nuclear Research \\
Dubna, 141980, Russian Federation}

\abstract{The dispersive approach to QCD is briefly overviewed and its
application to the assessment of hadronic contributions to electroweak
observables is discussed.}

\maketitle

Various strong interaction processes, as well as the hadronic
contributions to electroweak observables, are governed by the hadronic
vacuum polarization function~$\Pi(q^2)$, related Adler
function~$D(Q^2)$~\cite{Adler}, and the function~$R(s)$, which is
identified with the $R$--ratio of electron--positron annihilation into
hadrons. The hadronic vacuum polarization function constitutes the scalar
part of the hadronic vacuum polarization tensor
\begin{equation}
\label{P_Def}
\Pi_{\mu\nu}(q^2) = i\!\int\!d^4x\,e^{i q x} \bigl\langle 0 \bigl|
T\bigl\{J_{\mu}(x)\, J_{\nu}(0)\bigr\} \bigr| 0 \bigr\rangle =
\frac{i}{12\pi^2}\, (q_{\mu}q_{\nu} - g_{\mu\nu}q^2)\, \Pi(q^2),
\end{equation}
whereas the definition of functions~$R(s)$ and~$D(Q^2)$ is given in
Eqs.~(\ref{R_Def}) and~(\ref{Adler_Def}), respectively. The~QCD asymptotic
freedom makes it possible to study the high--energy behavior of the
functions~$\Pi(q^2)$ and~$D(Q^2)$ directly within perturbation theory.
At~the same time, the description of the function~$R(s)$ additionally
requires the use of pertinent dispersion relations, see
papers~\cite{Rad82, KP82, Pi2Terms1, Pi2Terms2, Pi2Terms3} and references
therein. As~for the low--energy hadron dynamics, it can only be accessed
within nonperturbative approaches, for instance, analytic
gauge--invariant~QCD~\cite{Fried1, Fried2, Fried3, Fried4},
holographic~QCD~\cite{HolQCD3, HolQCD4}, Schwinger--Dyson
equations~\cite{SDE1, SDE2, SDE3}, Bethe--Salpeter equations~\cite{BSE1,
BSE2, APT4, PRL99PRD77}, lattice simulations~\cite{LattRev, Lat1, Lat2,
Lat3, Lat4a, Lat4b}, operator product expansion~\cite{OPE1, OPE23, OPE4,
OPE5, OPE6}, nonlocal chiral quark model~\cite{NLCQM, ILM}, and others
(see, e.g., Ref.~\cite{Frasca}).

A~certain nonperturbative hint about the strong interactions in the
infrared domain is provided by the relevant dispersion relations. The
latter are widely employed in a variety of issues of contemporary
theoretical particle physics, such as, for example, the extension of
applicability range of chiral perturbation theory~\cite{Portoles,
Passemar}, the precise determination of parameters of
resonances~\cite{Kaminski}, the assessment of the hadronic
light--by--light scattering~\cite{DispHlbl}, and many others (see, e.g.,
papers~\cite{APT, APT1, APT2, APT3, PRD6264, Review, APT5a, APT5b,
Schrempp, APT6, APT7a, APT7b, APT8, APT9, APT10, APT11, BCmath} and
references therein).

Basically, the dispersion relations render the kinematic restrictions on
pertinent physical processes into the mathematical form and thereby impose
stringent intrinsically nonperturbative constraints on the relevant
quantities. In particular, the complete set of dispersion relations
(see Refs.~\cite{Adler, Rad82, KP82, Pivovarov91, PRD88})
\begin{equation}
\label{P_Disp}
\DP(q^2\!,\, q_0^2) = (q^2 - q_0^2) \int_{m^2}^{\infty}
\frac{R(\sigma)}{(\sigma-q^2)(\sigma-q_0^2)}\, d\sigma =
- \int_{-q_0^2}^{-q^2} D(\zeta) \frac{d \zeta}{\zeta},
\end{equation}
\begin{equation}
\label{R_Def}
R(s) = \frac{1}{2 \pi i} \lim_{\varepsilon \to 0_{+}}
\Bigl[\Pi(s + i \varepsilon) - \Pi(s - i \varepsilon)\Bigr] =
\frac{1}{2 \pi i} \lim_{\varepsilon \to 0_{+}}
\int_{s + i \varepsilon}^{s - i \varepsilon}
D(-\zeta)\,\frac{d \zeta}{\zeta},
\end{equation}
\begin{equation}
\label{Adler_Def}
D(Q^2) = - \frac{d\, \Pi(-Q^2)}{d \ln Q^2} =
Q^2 \int_{m^2}^{\infty} \frac{R(\sigma)}{(\sigma+Q^2)^2}\, d\sigma
\end{equation}
supplies substantial restraints on the infrared behavior of the functions
on hand, that, in turn, plays an essential role in the study of the
pertinent strong interaction processes at low energies, see
papers~\cite{DPT1a, PRD88, JPG42} for the details. In
Eqs.~(\ref{P_Disp})--(\ref{Adler_Def}) $Q^2 = -q^2 > 0$ and $s = q^2 > 0$
stand for the spacelike and timelike kinematic variables, respectively,
$m^2 = 4m_{\pi}^2$ is the hadronic production threshold, and $\DP(q^2\!,\,
q_0^2) = \Pi(q^2) - \Pi(q_0^2)$ denotes the subtracted hadronic vacuum
polarization function.

The dispersive approach to QCD~\cite{DPT1a, PRD88, JPG42} (its preliminary
formulation was discussed in Refs.~\cite{DPTPrelim1, DPTPrelim2}) merges
the aforementioned nonperturbative constraints with corresponding
perturbative input and provides the following unified integral
representations for the functions on hand:
\begin{eqnarray}
\label{P_DPT}
\DP(q^2,\, q_0^2) \eqnsgn{=} \DP^{(0)}(q^2,\, q_0^2) +
\int_{m^2}^{\infty} \rho(\sigma)
\ln\left(\frac{\sigma-q^2}{\sigma-q_0^2}
\frac{m^2-q_0^2}{m^2-q^2}\right)\frac{d\,\sigma}{\sigma}, \\
\label{R_DPT}
R(s) \eqnsgn{=} R^{(0)}(s) + \theta(s-m^2) \int_{s}^{\infty}\!
\rho(\sigma) \frac{d\,\sigma}{\sigma}, \\
\label{Adler_DPT}
D(Q^2) \eqnsgn{=} D^{(0)}(Q^2) +
\frac{Q^2}{Q^2+m^2}
\int_{m^2}^{\infty} \rho(\sigma)
\frac{\sigma-m^2}{\sigma+Q^2} \frac{d\,\sigma}{\sigma}.
\end{eqnarray}
In these equations $\theta(x)$~is the unit step--function [$\theta(x)=1$
if $x \ge 0$ and $\theta(x)=0$ otherwise], the leading--order terms
read~\cite{Feynman, QEDAB}
\begin{eqnarray}
\label{P0L}
\DP^{(0)}(q^2,\, q_0^2) \eqnsgn{=} 2\,\frac{\varphi - \tan\varphi}{\tan^3\varphi}
- 2\,\frac{\varphi_{0} - \tan\varphi_{0}}{\tan^3\varphi_{0}},
\qquad \sin^2\!\varphi = \frac{q^2}{m^2},
\qquad \sin^2\!\varphi_{0} = \frac{q^{2}_{0}}{m^2}, \\
\label{R0L}
R^{(0)}(s) \eqnsgn{=} \theta(s - m^2)\left(1-\frac{m^2}{s}\right)^{3/2}, \\
\label{D0L}
D^{(0)}(Q^2) \eqnsgn{=} 1 + \frac{3}{\xi}\left[1 - \sqrt{1 + \xi^{-1}}\,
\mbox{arcsinh}\,\bigl(\xi^{1/2}\bigr)\right],
\qquad \xi=\frac{Q^2}{m^2},
\end{eqnarray}
and $\rho(\sigma)$ denotes the spectral density
\begin{equation}
\label{RhoGen}
\rho(\sigma) = \frac{1}{\pi} \frac{d}{d\,\ln\sigma}\,
\mbox{Im}\lim_{\varepsilon \to 0_{+}} p(\sigma-i\varepsilon) =
- \frac{d\, r(\sigma)}{d\,\ln\sigma} =
\frac{1}{\pi}\,\rm{Im}\lim_{\varepsilon \to 0_{+}} {\it d}(-\sigma-i\varepsilon),
\end{equation}
with $p(q^2)$, $r(s)$, and~$d(Q^2)$ being the strong corrections to the
functions~$\Pi(q^2)$, $R(s)$, and~$D(Q^2)$, respectively, see
papers~\cite{DPT1a, PRD88, JPG42} and references therein for the details.
The~integral representations (\ref{P_DPT})--(\ref{Adler_DPT}) are by
construction consistent with the foregoing nonperturbative constraints and
constitute the ``dispersively improved perturbation theory''~(DPT)
expressions for the functions on hand.

The rigorous calculation of the spectral density~(\ref{RhoGen}), which
would have thoroughly accounted for both perturbative and nonperturbative
aspects of hadron dynamics, is a rather challenging and hardly feasible
(at least, at the present time) objective. Nonetheless, it appears that
even with such incomplete input as the perturbative part of the spectral
density
\begin{equation}
\label{RhoPert}
\rho\ind{}{pert}(\sigma) = \frac{1}{\pi} \frac{d}{d\,\ln\sigma}\,
\mbox{Im}\lim_{\varepsilon \to 0_{+}} p\ind{}{pert}(\sigma-i\varepsilon) =
- \frac{d\, r\ind{}{pert}(\sigma)}{d\,\ln\sigma}
= \frac{1}{\pi}\, \mbox{Im}\lim_{\varepsilon \to 0_{+}}
d\ind{}{pert}(-\sigma-i\varepsilon)
\end{equation}
the integral representations (\ref{P_DPT})--(\ref{Adler_DPT}) are capable
of yielding a physically sound behavior of the functions on hand in the
entire energy range. At~the one--loop level Eq.~(\ref{RhoPert}) assumes a
quite simple form, namely, $\rho\ind{(1)}{pert}(\sigma) = (4/\beta_{0})
[\ln^{2}(\sigma/\Lambda^2)+\pi^2]^{-1}$, where $\beta_{0} = 11 - 2\nf/3$,
$\nf$~denotes the number of active flavors, and~$\Lambda$ is the QCD scale
parameter. The explicit expressions for the spectral
function~(\ref{RhoPert}) up to the four--loop level are given in
Ref.~\cite{CPC} (recently calculated respective four--loop perturbative
coefficient can be found in Ref.~\cite{AdlerPert4L}). The perturbative
spectral function~(\ref{RhoPert}) will be employed hereinafter.

It has to be mentioned that in the massless limit $(m=0)$ for the case of
perturbative spectral function~(\ref{RhoPert}) the relations for the
function~$R(s)$~(\ref{R_DPT}) and the Adler
function~$D(Q^2)$~(\ref{Adler_DPT}) become identical to those of the
``analytic perturbation theory''~(APT)~\cite{APT} (see also
Refs.~\cite{APT1, APT2, APT3, PRD6264, Review, APT5a, APT5b, Schrempp,
APT6, APT7a, APT7b, APT8, APT9, APT10, APT11, BCmath}), whereas the
hadronic vacuum polarization function~$\Pi(q^2)$ was not addressed in the
framework of the latter. However, as discussed in, e.g.,
papers~\cite{PRD88, DPT1a, JPG42, C12, DPT2}, the massless limit ignores
substantial nonperturbative constraints, which relevant dispersion
relations impose on the functions on hand, that appears to be essential
for the studies of hadron dynamics at low energies.

\begin{figure}[t]
\centering
\includegraphics[width=75mm]{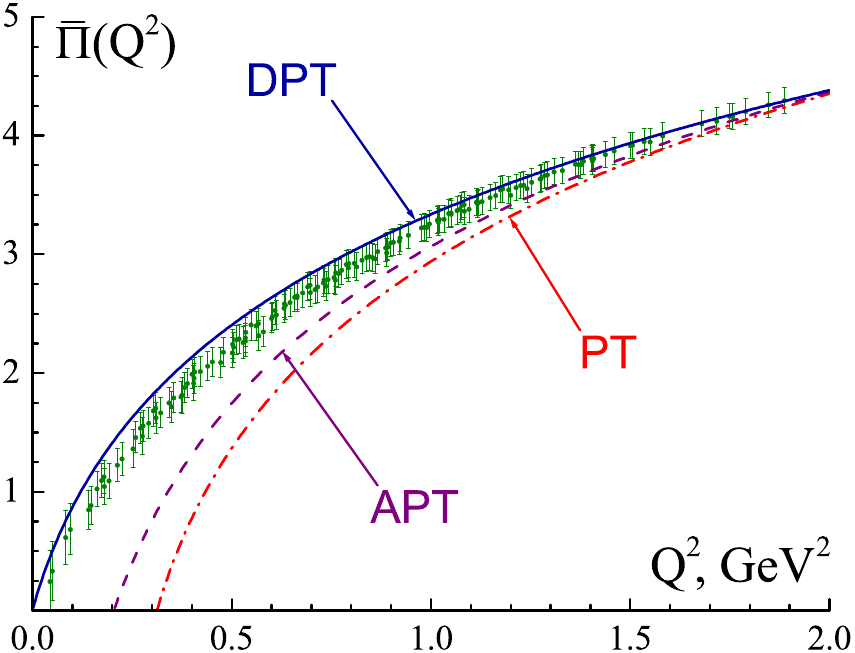}
\caption{Hadronic vacuum polarization function within various approaches:
dispersively improved perturbation theory~(label DPT, Eq.~(\ref{P_DPT}),
solid curve), its massless limit~(label APT, dashed curve), perturbative
approximation~(label PT, dot--dashed curve), and lattice simulation
data~(Ref.~\cite{Lat5}, circles).}
\label{Plot:P_DPT}
\end{figure}

The dispersive approach to QCD~\cite{DPT1a, PRD88, JPG42} enables one to
get rid of some inherent obstacles of the QCD perturbation theory and
significantly extends its range of applicability toward the infrared
domain. In~particular, the Adler function~(\ref{Adler_DPT}) agrees with
its experimental prediction in the entire energy range~\cite{DPT1a, DPT1b,
DPT2} (the studies of~$D(Q^2)$ within other approaches can be found in,
e.g., Refs.~\cite{MSS, Cvetic, Maxwell, Kataev, Fischer, PeRa, BJ}). The
representation~(\ref{Adler_DPT}) also complies with the results of
Bethe--Salpeter calculations~\cite{PRL99PRD77} as well as of lattice
simulations~\cite{RhoRescale1, RCTaylor}. The~dispersive approach to QCD
is capable of describing OPAL (update~2012, Ref.~\cite{OPAL9912}) and
ALEPH (update~2014, Ref.~\cite{ALEPH0514}) experimental data on inclusive
$\tau$~lepton hadronic decay in both vector and axial--vector channels in
a self--consistent way~\cite{PRD88, QCD14} (see also Refs.~\cite{DPT3,
C12}). Additionally, as one can infer from Fig.~\ref{Plot:P_DPT}, the DPT
expression for the hadronic vacuum polarization function
(Eq.~(\ref{P_DPT}), solid curve) is in a good agreement with lattice
simulation data~\cite{Lat5}~(circles) (the rescaling procedure described
in Refs.~\cite{RhoRescale1, RhoRescale2} was applied). The displayed in
Fig.~\ref{Plot:P_DPT} result corresponds to the four--loop level,
$\Lambda=419\,$MeV, and~$\nf=2$ active flavors. Figure~\ref{Plot:P_DPT}
also presents the massless limit of Eq.~(\ref{P_DPT}) (label~APT, dashed
curve) and the perturbative approximation of~$\Pi(q^2)$ (label~PT,
dot--dashed curve). As~one can infer from this figure, the perturbative
approximation of~$\Pi(q^2)$ contains infrared unphysical singularities,
that invalidates it at low energies, whereas the APT prediction diverges
in the infrared limit~$Q^2 \to 0$, that makes it also inapplicable at low
energies.

\bigskip

Let us address now the hadronic contributions to electroweak observables.
One of the most challenging issues of the elementary particle physics,
which engages the entire pattern of interactions within the Standard
Model, is the muon anomalous magnetic moment~\mbox{$a_{\mu} =
(g_{\mu}-2)/2$}. The persisting few standard deviations discrepancy
between the experimental measurements~\cite{MuonExp1, MuonExp2} and
theoretical evaluations~\cite{MuonRev1, MuonRev2} of this quantity may be
an evidence for the existence of a new physics beyond the Standard Model,
that brings the accuracy of an estimation of~$a_\mu$ to the top of the
agenda.

The uncertainty of theoretical estimation of~$a_{\mu}$ is mainly dominated
by the leading--order hadronic contribution~\cite{Raf72}
\begin{equation}
\label{AmuHVP}
a_{\mu}^{\mbox{\tiny HLO}} =
\frac{1}{3} \left(\frac{\alpha}{\pi}\right)^{\!2}
\!\int_{0}^{1}\!(1-x)\,
\bar{\Pi}\left(m_{\mu}^{2}\,\frac{x^2}{1-x}\right) dx, \qquad
\bar{\Pi}(Q^2) = \DP(0,-Q^2).
\end{equation}
This expression involves the integration of~$\Pi(q^2)$ over the
low--energy range, that basically gives the major contribution
to~$a_{\mu}^{\mbox{\tiny HLO}}$. However, the infrared behavior of the
hadronic vacuum polarization function is inaccessible within perturbation
theory. One of the ways to overcome this obstacle is to express the
integrand of Eq.~(\ref{AmuHVP}) in terms of the $R$--ratio of
electron--positron annihilation into hadrons and substitute its
low--energy behavior by the corresponding experimental measurements, see,
e.g., Refs.~\cite{HLMNT11, BDDJ15, DHMZ11}.

\begin{figure}[t]
\centering
{\includegraphics[width=75mm,clip]{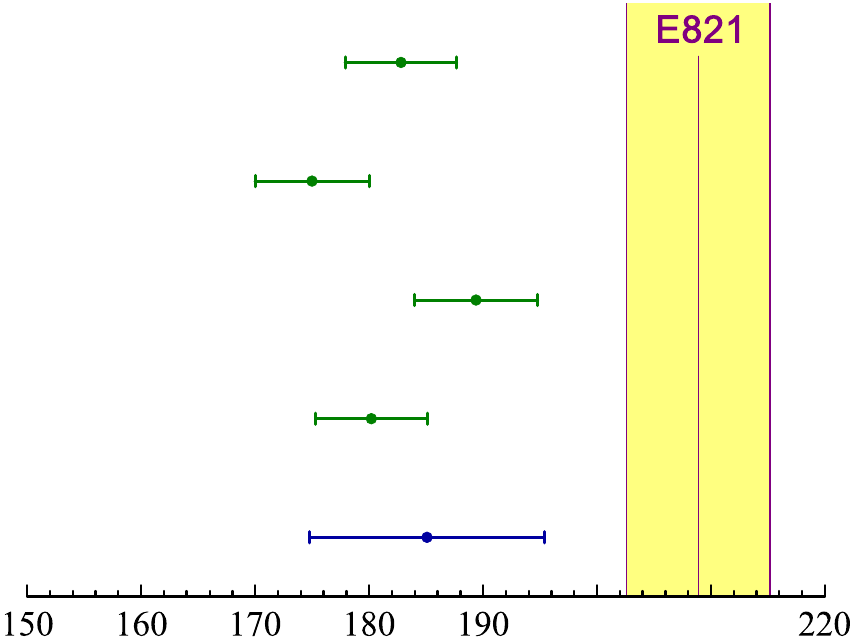}%
\hbox to 0pt {\hss%
\unitlength=1mm
\begin{picture}(74,53.5)
\put(52.075,-0.5){\small$\Delta a_{\mu}\!\times\! 10^{10}$}
\put(2.5,49.9){\txt{HLMNT'11~\cite{HLMNT11}}{G4}}
\put(2.5,39.5){\txt{BDDJ'15~\cite{BDDJ15}}{G4}}
\put(2.5,29.1){\txt{DHMZ'11($\tau$)~\cite{DHMZ11}}{G4}}
\put(2.5,18.7){\txt{DHMZ'11(e)~\cite{DHMZ11}}{G4}}
\put(2.5,8.3){\txt{This work}{RB4}}
\end{picture}}}
\unitlength=1pt
\caption{The subtracted muon anomalous magnetic moment ($\Delta a_{\mu} =
a_{\mu} - a_{0}$, $\,a_{0} = 11659 \times 10^{-7}$): theoretical
evaluations (circles) and experimental measurement (Eq.~(\ref{Amu_exp}),
shaded band).}
\label{Plot:Amu}
\end{figure}
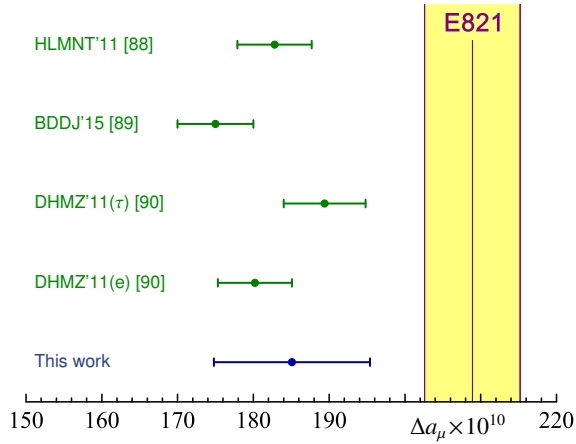

At the same time, as elucidated above, the DPT expression
for~$\Pi(q^2)$~(\ref{P_DPT}) contains no unphysical singularities, that,
in turn, enables one to perform the integration in Eq.~(\ref{AmuHVP}) in a
straightforward way, namely, without invoking experimental data
on~$R$--ratio. Eventually this results in the following assessment
\begin{equation}
\label{AmuHLO_DPT}
a_{\mu}^{\mbox{\tiny HLO}} = (696.1 \pm 9.5) \times 10^{-10},
\end{equation}
see Refs.~\cite{JPG42, HCEWDPT}. Equation~(\ref{AmuHLO_DPT}) corresponds
to the four--loop level and the quoted error accounts for the
uncertainties of the parameters entering Eq.~(\ref{AmuHVP}), their values
being taken from Ref.~\cite{PDG2012CODATA2012}. The obtained estimation
of~$a_{\mu}^{\mbox{\tiny HLO}}$~(\ref{AmuHLO_DPT}) proves to be in a good
agreement with its recent evaluations~\cite{HLMNT11, BDDJ15, DHMZ11}.

In addition to the leading--order hadronic
contribution~$a_{\mu}^{\mbox{\tiny HLO}}$~(\ref{AmuHLO_DPT}), the complete
muon anomalous magnetic moment~$a_{\mu}$ includes the
higher--order~\cite{HLMNT11} and light--by--light~\cite{AmuHlbl} hadronic
contributions, the QED contribution~\cite{AmuQED}, and the electroweak
contribution~\cite{AmuEW}, that altogether yields
\begin{equation}
\label{Amu_DPT}
a_{\mu} = (11659185.1 \pm 10.3) \times 10^{-10},
\end{equation}
see Refs.~\cite{JPG42, HCEWDPT}. The discrepancy between the experimental
value~\cite{MuonExp2, MuonExp3}
\begin{equation}
\label{Amu_exp}
a_{\mu}^{\mbox{\scriptsize exp}} = (11659208.9 \pm 6.3) \times 10^{-10}
\end{equation}
and the estimation~(\ref{Amu_DPT}) corresponds to two~standard deviations.
As one can infer from Fig.~\ref{Plot:Amu}, the obtained value of the muon
anomalous magnetic moment $a_{\mu}$~(\ref{Amu_DPT}) also complies with its
recent assessments~\cite{HLMNT11, BDDJ15, DHMZ11}.

\begin{figure}[t]
\centering
{\includegraphics[width=75mm,clip]{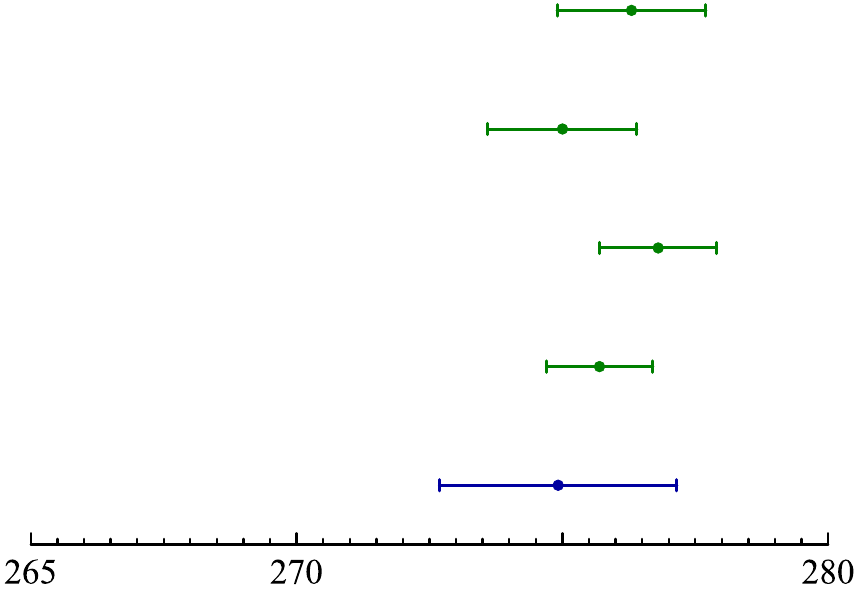}%
\hbox to 0pt {\hss%
\unitlength=1mm
\begin{picture}(74,53.5)
\put(40.5,-0.5){\small$\Delta\alpha\ind{(5)}{had}(M\inds{2}{Z})\!\times\! 10^{4}$}
\put(3,50.1){\txt{HLMNT'11~\cite{HLMNT11}}{G4}}
\put(3,39.7){\txt{J'11~\cite{J11}}{G4}}
\put(3,29.3){\txt{DHMZ'12($\tau$)~\cite{DHMZ11}}{G4}}
\put(3,18.9){\txt{DHMZ'12(e)~\cite{DHMZ11}}{G4}}
\put(3,8.5){\txt{This work}{RB4}}
\end{picture}}}
\unitlength=1pt
\caption{Theoretical evaluations of the five--flavor hadronic contribution
to the shift of electromagnetic fine structure constant at the scale of
$Z$~boson mass.}
\label{Plot:AHMZ}
\end{figure}
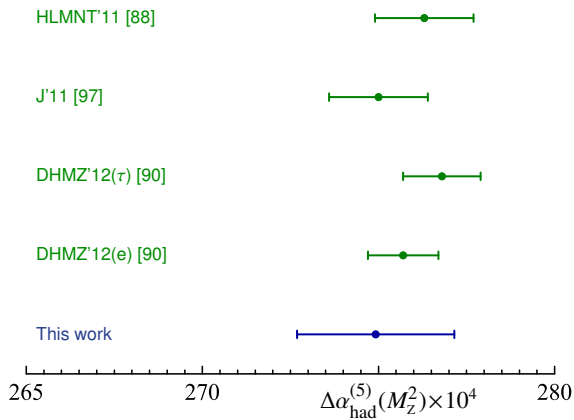

Another observable of an apparent interest is the electromagnetic running
coupling
\begin{equation}
\label{RC_QED}
\alpha\ind{}{em}(q^2) = \frac{\alpha}{1 - \AL{}(q^2) - \AH{}(q^2)},
\end{equation}
which plays a crucial role in a variety of issues of precision particle
physics. The leptonic contribution~$\AL{}(q^2)$ to Eq.~(\ref{RC_QED}) can
reliably be calculated by making use of perturbation theory~\cite{AEMLep}.
However, similarly to the earlier discussed case of the muon anomalous
magnetic moment, the hadronic contribution to Eq.~(\ref{RC_QED})
\begin{equation}
\label{AMH}
\AMH{}(q^2) = - \frac{\alpha}{3\pi}\,q^2
\,\mathcal{P}\!\!\int_{m^2}^{\infty}\!
\frac{R(s)}{s-q^2}\,\frac{d s}{s}
\end{equation}
involves the integration over the low--energy range and constitutes the
major source of the uncertainty in the assessment
of~$\alpha\ind{}{em}(q^2)$, see, e.g., Refs.~\cite{HLMNT11, Passera}.
In~Eq.~(\ref{AMH}) $\mathcal{P}$~stands for the ``Cauchy principal
value''.

The five--flavor hadronic contribution to the shift of the electromagnetic
fine structure constant at the scale of $Z$~boson mass can be evaluated in
the framework of~DPT in the very same way as above, that leads~to
\begin{equation}
\label{AMH_DPT}
\AMH{(5)}(M\inds{2}{Z}) = (274.9 \pm 2.2) \times 10^{-4},
\end{equation}
see Refs.~\cite{JPG42, HCEWDPT}. The estimation~(\ref{AMH_DPT})
corresponds to the four--loop level and the quoted error accounts for the
uncertainties of the parameters entering Eq.~(\ref{AMH}), their values
being taken from Ref.~\cite{PDG2012CODATA2012}. As one can infer from
Fig.~\ref{Plot:AHMZ}, the obtained assessment
of~$\AMH{(5)}(M\inds{2}{Z})$~(\ref{AMH_DPT}) proves to be in a good
agreement with its recent evaluations~\cite{HLMNT11, DHMZ11, J11}. The
corresponding value of the electromagnetic running
coupling~$\alpha\ind{}{em}(M\inds{2}{Z})$~(\ref{AMH}) additionally
includes leptonic~\cite{AEMLep} and top~quark~\cite{AEMtop} contributions,
that eventually yields
\begin{equation}
\label{AMZ_DPT}
\alpha\ind{-1}{em}(M\inds{2}{Z}) = 128.962 \pm 0.030.
\end{equation}
The obtained value~(\ref{AMZ_DPT}) also complies with recent assessments
of the quantity on hand~\cite{HLMNT11, DHMZ11, J11}, see
papers~\cite{JPG42, HCEWDPT} and references therein for the details.

\end{document}